\documentclass{emulateapj}

\newcommand{\dxdy}[2]{{\frac{\partial{#1}}{\partial{#2}}}}

\newcommand{\DxDy}

\shorttitle{Solar Rotation}
\shortauthors{Rogers}

\begin{document}


\title{On Limiting the Thickness of the Solar Tachocline}
  

\author{T.M. Rogers}
\affil{Department of Planetary Sciences, University of Arizona,
    Tucson, AZ, 85719}
\email{tami@lpl.arizona.edu}
\begin{abstract}

We present axisymmetric simulations of the coupled convective and radiative regions in the Sun in order to investigate the angular momentum evolution of the radiative interior. Both hydrodynamic and magnetohydrodynamic models were run.  We find an initial rapid adjustment in which the differential rotation of the convection zone viscously spreads into the radiative interior, thus forming a ``tachocline''.  In polar regions the subsequent spread of the tachocline is halted by a counter-rotating meridional circulation cell which develops in the tachocline.  Near the equator such a counter-rotating cell is more intermittent and the tachocline penetration depth continues to increase, albeit more slowly than previously predicted.   In the magnetic models we impose a dipolar field initially confined to the radiative interior.  The behavior of the magnetic models is very similar to their non-magnetic counterparts.  Despite being connected to the convection zone, very little angular momentum is transferred between the convective and radiative regions.  Therefore, while it appears that a magnetic field is not necessary to stop the tachocline spread, it also does not promote such a spread if connected to the convection zone.  

\end{abstract}


\section{Introduction}
Helioseismic observations \citep{brown89,goode91} have revealed a surprising picture of the solar rotation profile.  These observations show that the differential rotation observed at the surface, with the equator spinning faster than the poles, persists to the base of the convection zone with very little radial variation.  On the other hand the radiative interior is revealed to be rotating uniformly.  The transition between these very different rotation profiles is now commonly known as the tachocline, with a still undetermined extent, although \cite{elliott99} estimate a width of 0.019$R_{\odot}$.  

The observed rotation profile is puzzling for many reasons.  First, it was previously thought that the rotation profile of the solar convection zone would obey the Taylor-Proudman (TP) constraint, in which rotation is constant on cylinders.  Indeed, it has proven quite difficult to break this constraint and recover differential rotation constant on cones, similar to the observed profile.  \cite{re05} showed that the TP constraint could be overcome with a latitudinal entropy gradient imposed in a subadiabatic tachocline.  This thermal wind would break the TP constraint and lead to more realistic differential rotation in the convection zone.  Follow up numerical experiments by \cite{mie06} showed that indeed, when a latitudinal temperature gradient of order 10K is imposed as a boundary condition at the base of their model convection zone, solar-like differential rotation is recovered.  

The studies of the convection zone differential rotation discussed above suggest the importance of the underlying tachocline and adjacent stable region in understanding the differential rotation of the convection zone.  This leads to the second (and third) puzzling aspect of the solar rotation profile.  Despite the constant forcing by the overlying differential rotation of the convection zone the tachocline remains thin and the radiative interior is maintained in uniform rotation.  Although a thin tachocline does not guarantee a uniformly rotating radiative interior, the two questions have often been treated as one.  In the defining work on the tachocline, \cite{sz92} attribute the thinness of the tachocline to anisotropic turbulence.  They argue that, similar to the Earth's atmosphere, the solar tachocline, lying mostly in the stably stratified interior, opposes vertical turbulence.  They show that the width of the tachocline is then determined by the degree of anisotropy in the turbulence.  Simulations carried out by \cite{elliott97} confirmed this trend.

An alternative solution was offered by \cite{gm98}.  They argue that there are many examples in the Earth's atmosphere in which two-dimensional turbulence of the sort envisioned by \cite{sz92} conserve potential vorticity and thus drive a system {\it away} from uniform rotation.  These authors take the helioseismic results as an indication that there {\it must} exist a large scale magnetic field in the radiative interior.  Such a field would easily establish uniform rotation of the radiative interior via Ferraro's isorotation law \citep{ferraro} and a thin tachocline could be formed as something of a magnetic boundary layer.  

The magnetic solution to the rotation profile of the radiative interior was not unique to \cite{gm98}, it was investigated by other authors \citep{charbmac93,rk97,mc99}.  In particular \cite{mc99} elucidated a problem with this theory that is now known as the ``confinement'' problem.  That is, magnetic field naturally diffuses and any remnant field in the radiative interior would diffuse into the convection zone.  Once the field is in contact with the differentially rotating convection zone, a toroidal field and associated Lorentz force act to spread the differential rotation of the convection zone into the radiative interior.  The contribution of \cite{gm98} was to suggest that this diffusion could be controlled by meridional circulation driven in the convection zone and burrowing into the tachcoline; effectively holding down the magnetic field.  Subsequent studies on the ability of a meridional flow to confine the tachocline have produced mixed results: \cite{gar02,kr06} initially showed that such confinement by meridional circulation was possible.  Time-dependent, three-dimensional (3D) simulations by \cite{bz06} indicated that that it was not.  The differences in these works amounted to how deep the meridional circulation, driven in the convection zone, could penetrate into the radiative interior, and with what strength.  \cite{kr06} estimated that a flow speed of $10^{3}$ cm/s penetrating $10^{8}$cm would be sufficient to confine the field.  
 
The strength of meridional circulation and it's ability to penetrate the stable tachocline depends, in turn, on how the boundary condition with the convection zone is treated.  In all of these studies only the radiative interior was modeled, with the convection zone treated purely as a boundary condition which imposed the differential rotation.  \cite{gb08} showed that the depth of penetration of meridional circulation into the radiative interior and its strength depended sensitively on the boundary conditions assumed.  In reality, the convection zone imposes a complex boundary condition with (at least) large scale inflows and outflows, penetration of plumes and pumping of magnetic field.  However, coupling the convection zone and radiative interior is a difficult undertaking because of the vast discrepancy in the governing timescales for the convective and radiative zones in the Sun.  Here we report on axisymmetric, time-dependent simulations which solve self-consistently both the convection zone and the underlying radiative region.  

\section{Numerical Model}
\subsection{Equations}
The numerical method we employ here is identical to that described in \cite{ro11}, here we give a brief review and refer the reader there for more details.  We solve the full set of axisymmetric MHD equations in the anelastic approximation: 
\begin{equation}
\nabla\cdot(\overline{\rho}{\bf u})=0.
\end{equation}
\begin{equation}
\nabla\cdot{\bf B}=0
\end{equation}
\begin{eqnarray}
\lefteqn{\dxdy{{\bf u}}{t}+({\bf u}\cdot\nabla){\bf u}=-\nabla P - C\overline{g}\hat{r}+2({\bf u}\times\Omega)}\nonumber\\&
&{+\frac{1}{\rho}({\bf J}\times{\bf B})+\overline\nu(\nabla^{2}{\bf u}+\frac{1}{3}\nabla(\nabla\cdot{\bf u}))}
\end{eqnarray}
\begin{eqnarray}
\lefteqn{\dxdy{T}{t}+({\bf v}\cdot\nabla){T}=-u_{r}(\frac{d\overline{T}}{dr}-(\gamma-1)\overline{T}h_{\rho})}\nonumber\\&  
& {+(\gamma-1)Th_{\rho}u_{r}+\gamma\overline{\kappa}[\nabla^{2}T+(h_{\rho}+h_{\kappa})\dxdy{T}{r}]}
\end{eqnarray}
\begin{equation}
\dxdy{{\bf B}}{t}=\nabla\times({\bf u}\times{\bf B})+\eta\nabla^{2}{\bf B}
\end{equation}
Equations 1 and 2 ensure the mass and magnetic flux are conserved. In Equation 3, the momentum equation, C denotes the co-density \citep{bra95,rg05}, g the gravity, $\Omega$ the rotation rate, $\nu$ and $\kappa$ the viscous and thermal diffusivity respectively.  The inverse density and thermal diffusivity scale heights are denoted by $h_{\rho}$ and $h_{\kappa}$, $\gamma$ is the adiabatic index and all other variables take their usual meanings.  In all of the above equations reference state variables are denoted by an overbar.  Equation 4 reresents the energy equation written as a temperature equation, where $u_{r}$ represents the radial velocity.  The first term on the right hand side (rhs) of equation 4 represents the difference between the reference state temperature gradient and the adiabatic temperature gradient and allows us to represent strongly subadiabatic regions in addition to convection zones.  Equation 5 is the magnetic induction equation, in which $\eta$ represents the magnetic diffusivity.  The diffusivities have the following forms:
\begin{equation}
\begin{array}{rrr}
\kappa&=&\kappa_{m}\frac{16\sigma \overline{T}^{3}}{3\overline{\rho}^{2}kc_{p}}\\
\nu&=&\frac{\nu_{m}}{\rho}\\
\eta&=&const\\
\end{array}
\end{equation}
where $\kappa_{m}$ is $10^{6}$, $\nu_{m}$ is $5\times 10^{11}$ and $\eta$ is $5\times 10^{12}$, giving units of $cm^{2}/s$.  Values at the base of the convection zone are shown in Table 1.  It is worth mentioning here that because our thermal diffusivity, $\kappa$ is $10^{6}$ larger than solar the flux through the simulation is $10^{6}\times$ solar.  Using Mixing Length Theory (MLT) this would imply velocities which are $(10^{6})^{1/3}\approx100\times$ solar.

These equations are solved in axisymmetric spherical coordinates using a spherical harmonic decomposition in latitude and a finite difference scheme in the radial direction.  The maximum spherical harmonic degree, {\it l$_{max}$}, is 340, so that the grid resolution is 512 zones in latitude.  The radial resolution is 1500 zones, with 600 zones dedicated to the radiative interior and 900 zones covering the convection zone.  We solve the above equations from 0.15 R$_{\odot}$ to 0.95 R$_{\odot}$.  The radiative region extends from 0.15 R$_{\odot}$ to 0.71 R$_{\odot}$ and the convection zone extends from 0.71 R$_{\odot}$ to 0.90 R$_{\odot}$, for numerical reasons we impose an additional stably stratified region from 0.90 R$_{\odot}$ to 0.95 R$_{\odot}$.  The reference state thermodynamic variables are taken from a polynomial fit to the standard solar model from \cite{jcd96}.  We should note that because of the convection and the gravity waves in the deep radiative interior our timestep is severely limited and therefore, we are unable to run these simulations very long.  Our average timestep is approximately 10s and all of these simulations have run $\approx 2 \times 10^8$s.
 
\begin{table}
\begin{center}
\begin{tabular}{|*{4}{c|}} 
   \hline 
   Parameter & Symbol & Sun & Simulation  \\ 
   \hline
  thermal diffusivity & $\kappa$ & $10^7$ & $1.1 \times 10^{13}$ \\
  magnetic diffusivity & $\eta$ & $10^3$ & $5 \times 10^{12}$ \\
  viscous diffusivity & $\nu$ & $30$ & $2.5 \times 10^{12}$ \\
  \hline  
  Prandtl number & $\nu/\kappa$ & $3\times 10^{-6}$ & $ 0.22$ \\
  magnetic Prandtl nb & $\nu/\eta$ & $3\times 10^{-2}$ & $ 0.5$ \\
  Froude number & $(\Omega / N)^2$ & $2 \times 10^{-6}$ & $ 2\times10^{-6}$ \\
  Ekman number & $\!\nu /2\Omega R^2\!$ & $2 \times 10^{-15}$ & $ 2 \times 10^{-4}$ \\
  \hline 
  rotation frequency & $\Omega$ & $ 2.7 \times 10^{-6}$ & $ 2.7 \times 10^{-6}$ \\ 
  Ohmic diffusion time & $t_{B}$ & $2.5 \times 10^{18}$ & $ 5.0 \times 10^{8}\!$ \\
  viscous diffusion time & $t_{\nu}$ & $\!8.333 \times 10^{19}\!$ & $ 10^{9}\!$ \\
  Eddington-Sweet time & $t_{ES}$ & $2.8 \times 10^{19}$ & $ 2.8 \times 10^{13}\!$ \\
   \hline 
\end{tabular}
\bigskip 
  \caption{Values for various parameters in the simulations.  The values listed for the diffusivities are those at the top of the radiation zone and have units $cm^{2}/s$.  Likewise, the timescales quoted are simply calculated using the diffusivities at the top of the radiative interior.  Diffusion times are calculated using the depth of the radiation zone and are quoted in seconds.  Each model was run approximately $2 \times 10^{8}s$ and required approximately a few hundred thousand processor hours.}   
 \label{tble}
   \end{center} 
\end{table}

\subsection{Model Setup}

We recognize that without fully 3D simulations we would be unable to recover the observed solar differential rotation profile in the convection zone.  However, we are mainly interested in studying the response of the underlying radiative interior to this differential rotation in the presence of physical phenomena such as convective overshoot and magnetic pumping.  Therefore, as in \cite{ro11} we artificially force the differential rotation of the convection zone to be that inferred from helioseismology \citep{thomp96} and described by: 
\begin{equation}
\Omega_{cz}=A+B\cos^{2}\theta+C\cos^{4}\theta
\end{equation}
where A is 456nHz, B is -72nHz and C is -42nHz.  Unlike \cite{ro11} we force the differential rotation only in the convection zone, with a step function to the uniform rotation of the radiative interior.\footnote{While we have run models both with an imposed tachocline and those without, we feel that imposing a tachocline does not accurately represent the situation we are trying to investigate.  This is because if a tachocline is forced, there is no stress presented to the radiative interior.}  The differential rotation is imposed through a forcing term on the azimuthal component of the momentum equation, {\it which is present only in the convection zone}: 

\begin{eqnarray}
  \dxdy{}{t}\left(\frac{u_{\phi}}{r \sin\theta}\right)
  &=&-\frac{(\nabla\cdot{\bf u}{u_{\phi}})_{\phi}}{r\sin\theta}
  +\frac{[(\nabla\times{\bf B})\times{\bf B}]_{\phi}}{\bar{\rho}\mu r\sin\theta}+\frac{(2{\bf u}\times\Omega)_{\phi}}{r\sin\theta}\nonumber\\&
  &\mbox{}+\frac{\nu}{r\sin\theta}\nabla^{2}u_{\phi}+\frac{\nu u_{\phi}}{r^{3}\sin^{3}\theta}+F
\end{eqnarray}
where the last term, F, represents the forcing function, defined as:
\begin{equation}
\begin{array}{rclr}
F&=&\frac{1}{\tau}\left(\Omega'(r,\theta)-\frac{u_{\phi}(r,\theta)}{r\sin\theta}\right) & \mbox{for $r > 0.71R_{\odot}$}\\
F&=&0. &\mbox{for $r < 0.71R_{\odot}$}\\
\end{array}
\end{equation}
where $\Omega'(r,\theta)$ represents the observed differential rotation relative to the constant $\Omega$  and $\tau$ is a forcing timescale.  In all simulations $\Omega$=441nHz and $\tau$ is $10^{4}$s.\footnote{We have run models with $\tau=10^{5}$ and $10^{3}$ for limited time (not as long as the full simulations presented here, but much longer than the forcing time) and see very little difference.}

\begin{figure}

%\epsscale{1.00}
%\plotone{f1.pdf}
\centering
\includegraphics[width=6in]{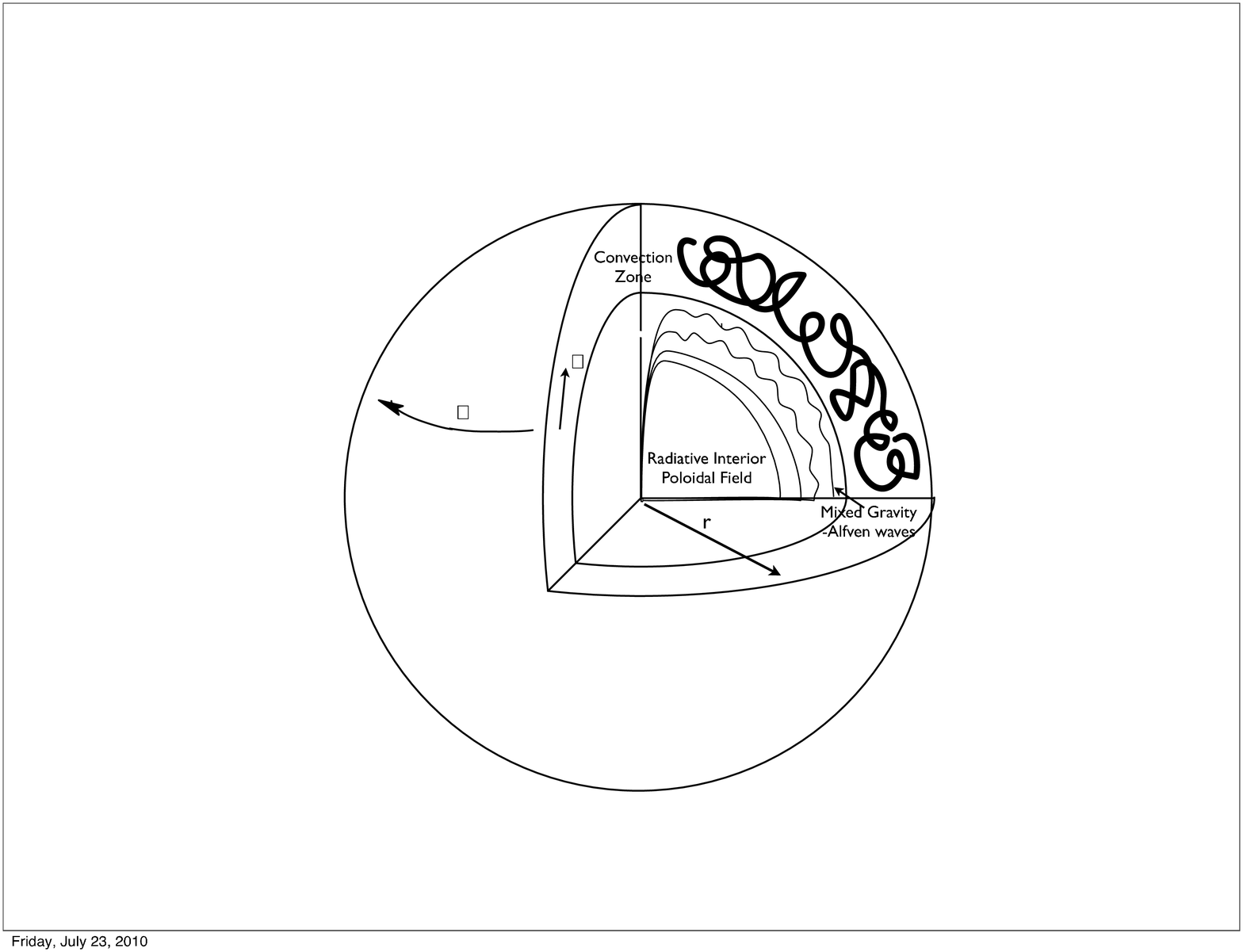}
\caption{Model schematic (reproduced from \cite{ro11}).  Radiation zone occupies the inner 75\% of the simulated domain, with convection occupying the outer 25\%.  In the magnetic models a dipolar field is imposed in the radiative interior (field lines shown).  Gravity-Alfven waves are generated in the radiative interior by impinging plumes.}
\label{fig:setup}
\end{figure}

We ran three models.  One purely hydrodynamic model and two magnetic models.  In the magnetic models we start with a hydrodynamic model until convection is sufficiently established at which point we imposed a dipolar field with the form:

\begin{equation}
B_{1}=B_{o}r^{2}(1-\frac{r}{0.71R_{\odot}})^{2}
\end{equation}
so that initially all field lines close just beneath the convection zone.  Figure~\ref{fig:setup} shows a schematic of the initial setup and model, reproduced from \cite{ro11}.  Although we ran models with two magnetic field strengths, $B_{o}=40G$ and $B_{o}=4000G$, all of the results presented will be only for the strong field case because the weak field case is virtually identical to the non-magnetic case.  

In all models the boundary conditions are stress-free and impermeable at the top and bottom boundaries.  The temperature boundary conditions are constant temperature perturbation on top and bottom and the magnetic field is matched to an internal potential field at the bottom of the domain and an external potential field at the top of the domain.  The current density, $J_{l}$ is zero on the top and bottom boundary.  

\section{Results}
\subsection{Meridional Circulation}

The fundamental assumption of the \cite{gm98} model is that meridional circulation driven by the differential rotation in the convection zone is able to penetrate (at least minimally) the radiative interior and confine the magnetic field.  One of the persistent issues in testing this model has been regarding how deep this meridional circulation penetrates the radiative interior, if at all.  \cite{gim04} argued that the depth of penetration of the meridional circulation driven within the convection zone was limited to an Ekman depth and therefore, would penetrate no further than the overshoot layer.  \cite{gb08} showed that the circulation driven beneath the convection zone depends strongly on the boundary conditions imposed at the top of the radiative layer and that radial forcing at the interface could produce small, but sustained flows deep within the interior.     

The simulations presented here couple the convective and radiative regions.  Moreover, the radiative interior is sufficiently ``stiff'', using the Brunt-Vaisala frequency from the standard solar model.  Therefore, these simulations are able to address the penetration of meridional circulation in the most self-consistent manner to date.  Furthermore, because we artificially impose the differential rotation in the convection zone we expect a faithful representation of the meridional circulation driven in the convection zone from the differential rotation.

\begin{figure}
%\epsscale{0.80}
%\plotone{f2.pdf}
\centering
\includegraphics[width=5in]{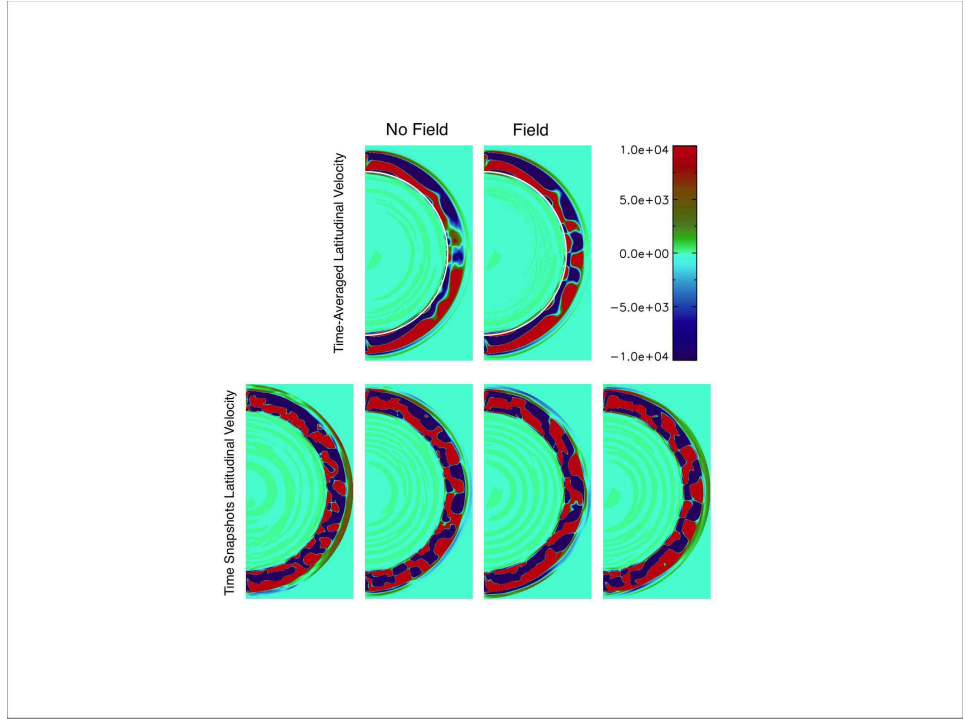}
\caption{Time averaged latitudinal flow in non-magnetic (a) and magnetic (b) models.  Red represents flow towards the south pole, while blue represents flow towards the north pole.  The white overlaid line in (a) and (b) represent the base of the convection zone.  Flow amplitudes are approximately $9\times 10^{4}$ cm/s at the top of the convection zone and $5\times 10^{4}$ cm/s at the base of the convection zone (colorbar units are in cm/s).  Bottom four plots show time snapshots from the non-magnetic model with time increasing to the right (c) $10^{4}$s, (d) $10^{7}$s, (e) $9\times 10^{7}$s and (f) $1.5\times 10^{8}$s.  Gravity waves are seen quite clearly in both time snapshots and time averages.}
\label{fig:latvel}
\end{figure}

\begin{figure}
\centering
%\plotone{f3.pdf}
%\epsscale{0.80}
\includegraphics[width=5in]{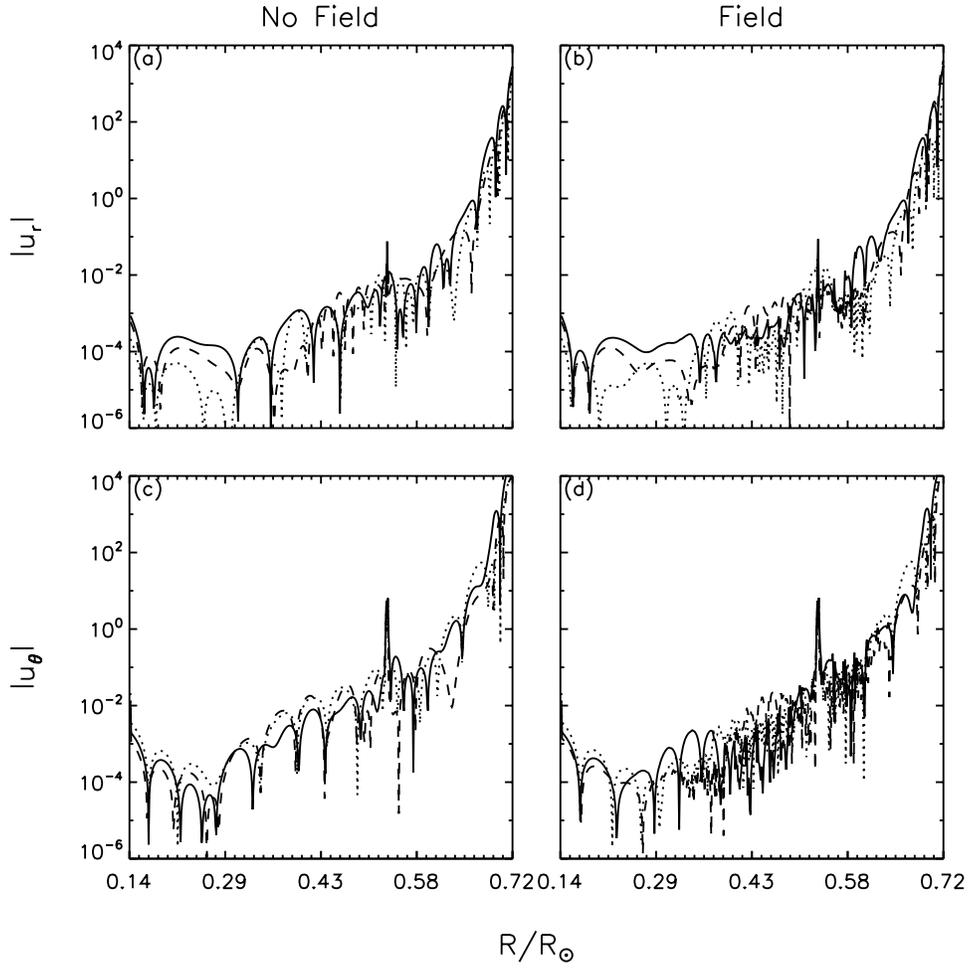}
\caption{Absolute magnitude of the time-averaged radial and latitudinal velocities in non-magnetic (a) and (c) and magnetic (b) and (d) models in cm/s.  Different line types represent different latitudes: 70$^{\circ}$ (solid line), 40$^{\circ}$ (dotted line) and 10$^{\circ}$ (dashed) lines.  We clearly see exponential decay of the amplitudes as well as the signs of gravity waves. Amplitudes in magnetic models show slightly faster decay and waves have shorter vertical wavelengths.}
\label{fig:velrad}
\end{figure}

In Figure~\ref{fig:latvel} we show the latitudinal velocity, both in time-average and time snapshots.  In this figure red represents flow towards the south pole, while blue represents flow towards the north pole.  One can immediately see the poleward flow at the surface, with equatorward flow near the base of the convection zone (the overlaid red line in (a) and (b) shows the base of the convection zone).  In both the magnetic and non-magnetic cases the time-averaged meridional circulation shows more structure near the equator and is not well described as a single cell, but this is likely due to needing longer time averages.  The amplitudes of the meridional circulation within the convection zone show velocities of approximately $9\times 10^{4}$ cm/s at the top of the convection zone and $5\times 10^{4}$ cm/s near the base of the convection zone.\footnote{The amplitudes of the meridional circulation here are larger than those observed in the solar convection zone because of an increased thermal diffusivity, as discussed in Section 2.1.  While this is not ideal, our overestimate of the flow speeds likely leads us to enhanced penetration of flows.  Weaker flows would lead to less penetration.}  One can also see hints of internal gravity waves (IGW), particularly in the non-magnetic case and in the time-snapshots (these can be seen more clearly in Figure~\ref{fig:velrad}).  

While Figure~\ref{fig:latvel} clearly shows the general circulation set up in the convection zone it is hard to discern the amplitudes of the meridional flows within the radiative interior.  In Figure~\ref{fig:velrad} we show these flows more clearly.  In that figure we show the absolute magnitude of the both the radial and latitudinal velocities as a function of radius below the convection zone for three different latitudes.  Very little latitudinal dependence is seen.  However, one can clearly see a rapid decrease in depth of the flow velocities, which are down nearly four to five orders of magnitude by 0.60$R_{\odot}$.  Figure~\ref{fig:velrad}a and \ref{fig:velrad}c show this amplitude for non-magnetic cases while \ref{fig:velrad}b and \ref{fig:velrad}d show it for the magnetic cases.  One can see a slightly faster decay in the magnetic cases as well as shorter vertical wavelengths of the mixed gravity-alfven waves, which is expected given the dispersion relation for these mixed waves \citep{macro11}.  

According to the analytic expression derived by \cite{gaa09}, the amplitude of the radial velocity should fall off exponentially with a length-scale determined by the parameter $\sigma=\sqrt{Pr}N/\Omega$, where N is the Brunt-Vaisala frequency.  In these simulations the Prandtl number and Brunt-Vaisala frequency, N, vary substantially over the domain, so a direct comparison is difficult, but using $N \approx 100 \mu Hz$ and a Prandtl number of 0.05 (the values just beneath the convection zone), the analytically predicted length-scale is $\approx 9 \times 10^{8}$cm.  Investigation of the data plotted in Figure~\ref{fig:velrad} gives a length-scale of $\approx 4 \times 10^{8}$cm, in fair agreement with the analytic prediction (given the simplified comparison).

\begin{figure}
\centering
%\plotone{f4.pdf}
%\epsscale{0.80}
\includegraphics[width=5in]{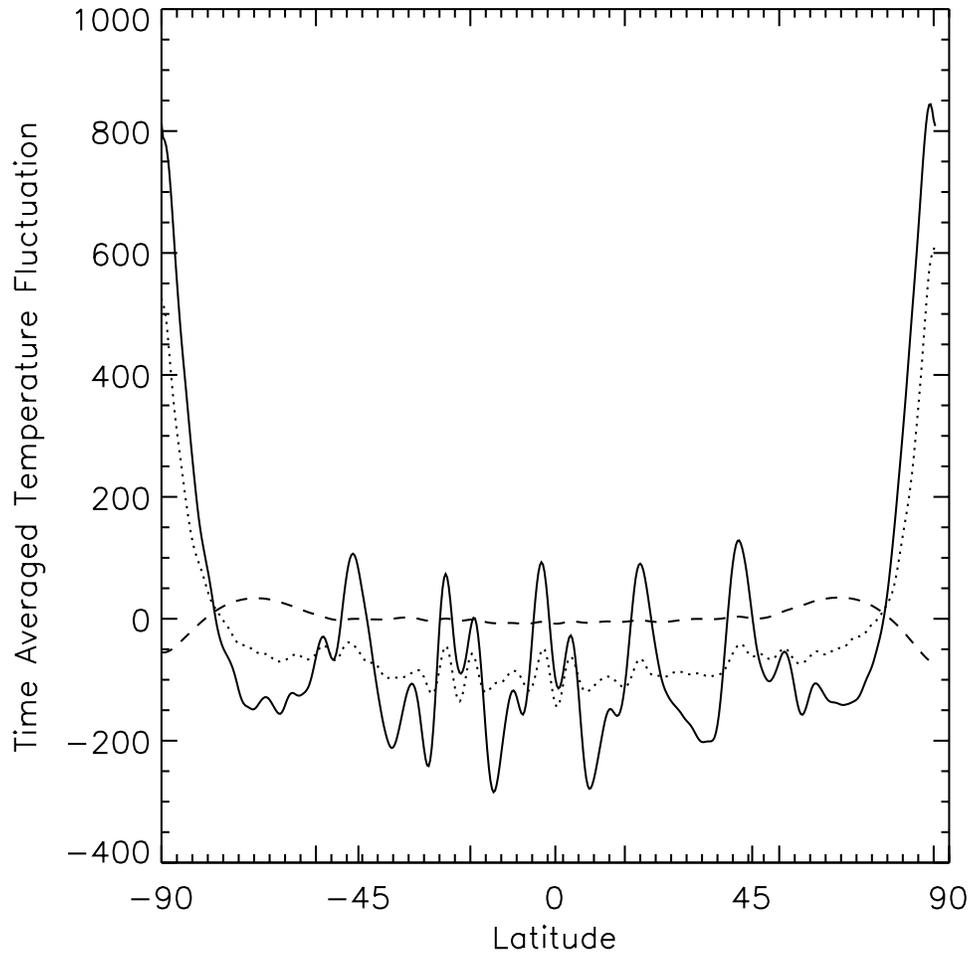}
\caption{Time averaged temperature variation as a function of latitude at $0.004 R_{\odot}$ ($3\times 10^{8}$cm) below the convection zone (solid line), $0.01 R_{\odot}$ ($8\times10^{8}$cm) below the convection zone (dotted line) and $0.03 R_{\odot}$ ($2\times10^{9}$cm) below the convection zone in Kelvin.  We clearly see warmer poles and cooler equator closer to the convection zone, although there is substantial variation at low latitudes.  Further from the convection zone this profile reverses, with slightly cooler poles and warmer equator.}
\label{fig:temfl}
\end{figure}

By forcing the differential rotation in the convection zone we are artificially breaking the Taylor-Proudman constraint there.  As predicted by \cite{re05} and tested by \cite{mie06} in three-dimensional simulations, breaking the Taylor-Proudman constraint could be achieved by a latitudinal temperature gradient within the tachocline (or as a bottom boundary condition in the case of the simulations).  Here we can ask the question in reverse; given the imposed differential rotation, is a latitudinal temperature gradient recovered?  In Figure~\ref{fig:temfl} we show the temperature fluctuation as a function of latitude at three radii within the tachocline (the tachocline will be defined in the next section, see figure caption for actual depths beneath the convection zone).  Here we see poles which are much warmer than the equator at the top of the tachocline (solid line) and slightly lower in the tachocline (dotted line).  However, we also see a fair amount of variation at lower latitudes.  In addition, near the middle of the tachocline (dashed line) this temperature variation in latitude switches sign (particularly at high latitudes), so that the poles are slightly cooler than the equator.  

Investigation of Figure~\ref{fig:latvel} indicates that the middle tachocline (where the latitudinal temperature gradient changes sign) is also assocated with a meridional circulation cell which rotates counter to the main meridional circulation driven in the convection zone.  This counter rotating cell (CRC) is robust in the polar regions.  Near the equator the cell is more time dependent, but longer time averages show the CRC becoming more prevalent.  We should note that such a CRC has been seen in 3D simulations by \cite{mie10}, however in that work they attribute this cell to their artificial impermeable boundary at the base of their convection zone.  We confirm here the presence of such a cell even under realistic boundary conditions.  This CRC has important implications for the spread of the tachocline as we will discuss in Section 3.4.

\subsection{Tachocline Penetration Depth}

\begin{figure}
%\epsscale{0.80}
%\plotone{f2.pdf}
\centering
\includegraphics[width=5in]{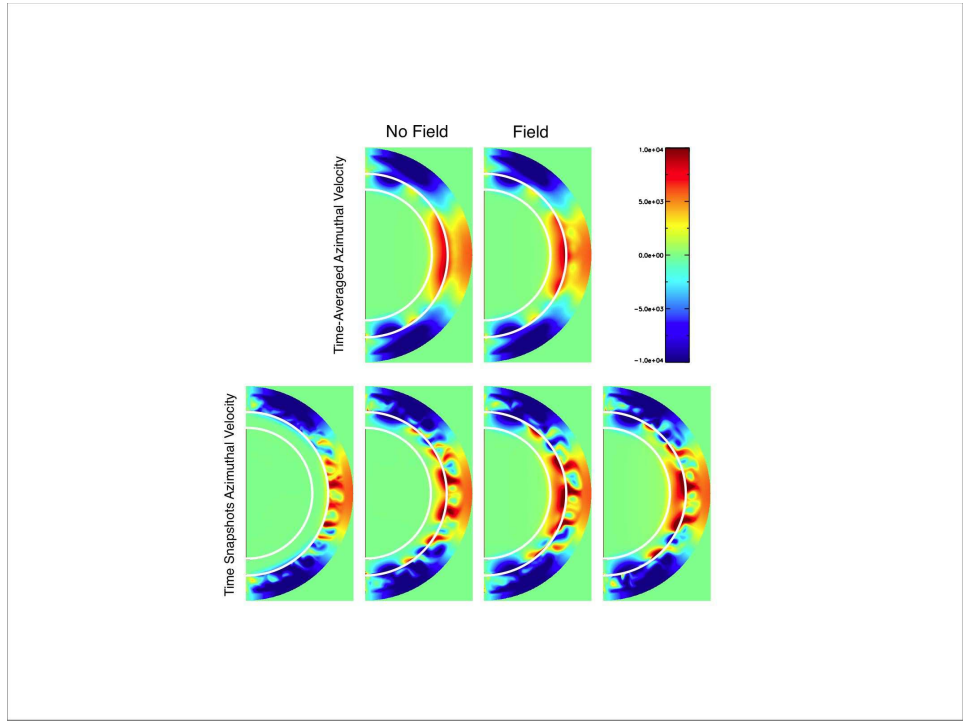}
\caption{Time averaged azimuthal flow in non-magnetic (a) and magnetic (b) models.  Red represents positive values with respect to the mean $\Omega$ (441nHz), while blue represents negative values with respect to the mean.  The overlaid white lines represent the base of the convection zone and the base of the tachocline after it has stopped spreading, or 0.14$R_{\odot}$ below the convection zone.  The bottom four plots show time snapshots from the non-magnetic model with time increasing to the right (c) $10^{4}$s, (d) $10^{7}$s, (e) $9\times 10^{7}$s and (f) $1.5\times 10^{8}$s.}
\label{fig:azvel}
\end{figure}

\begin{figure}
\centering
%\plotone{f5.pdf}
%\epsscale{0.80}
\includegraphics[width=5in]{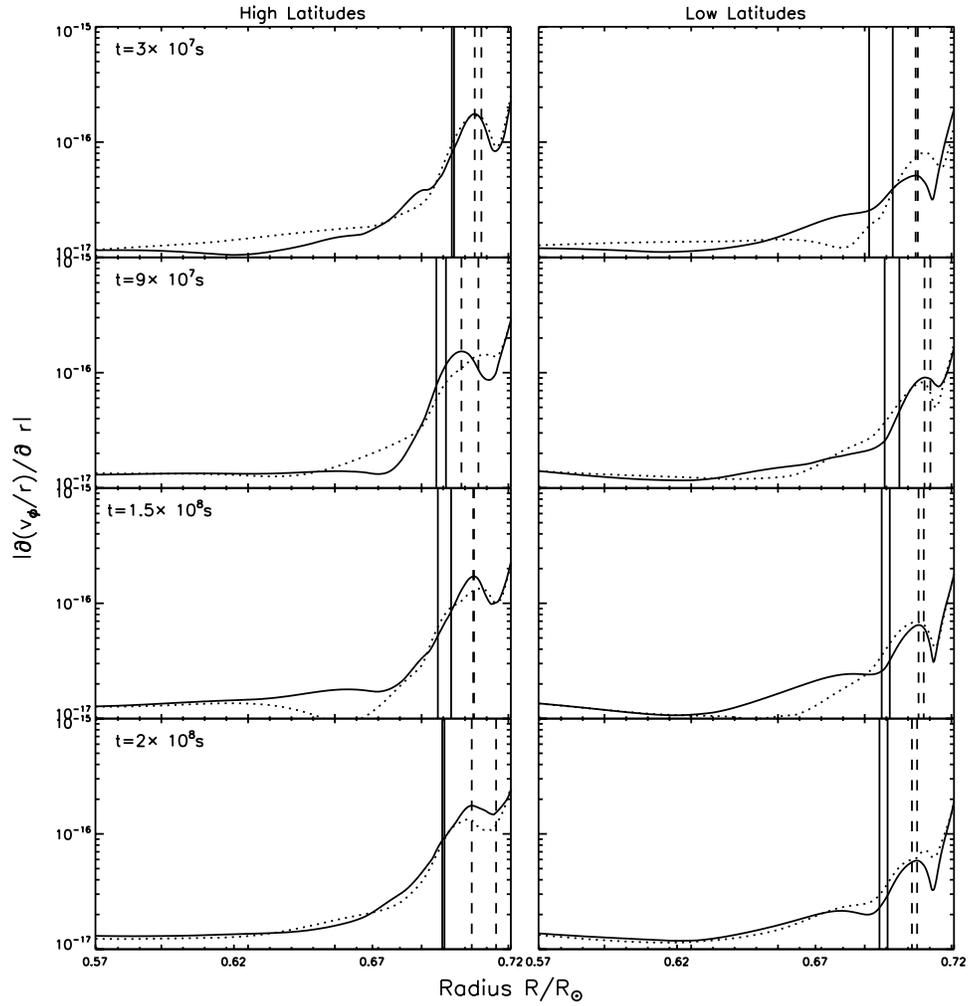}
\caption{Absolute value of the radial gradient of the angular velocity as a function of radius beneath the convection zone for the non-magnetic model (solid line) and the magnetic model (dotted line), in units of $1/cm/s$.  The low latitude values are averaged from 60$^{\circ}$-120$^{\circ}$ and the high latitude values are averaged from 0$^{\circ}$-30$^{\circ}$ and 150$^{\circ}$-180$^{\circ}$.  The half-maximum value is defined as half of the amplitude at the secondary peak (denoted by the dashed line) and marked with the solid line.  Defining the tachocline depth as distance between the base of the convection zone and this position, the tachocline penetration depth is measured as 0.021$R_{\odot}$ for high latitudes and 0.022$R_{\odot}$ for low latitudes in the non-magnetic model.}
\label{fig:fwhm}
\end{figure}

In the simulations discussed here we do not impose a tachocline, but instead want to investigate the spread of differential rotation from the convection zone into the radiative interior.  Thus, look at the formation of the tachocline and its evolution in time.  Figure~\ref{fig:azvel} shows both time-averaged and time snapshots of the azimuthal velocity.  One can clearly see that the differential rotation quickly penetrates into the radiative interior.  However, between Figure~\ref{fig:azvel}e and \ref{fig:azvel}f, there is very little change in that penetration, although panel ~\ref{fig:azvel}f represents a substantial increase in time.  One way to measure the tachocline depth is by looking at the region of strongest radial shear in the azimuthal velocity.  Looking at $|\partial{(v_{\phi}/r)}/\partial{r}|$ as a function of radius and time we can then measure the tahcocline penetration depth as the depth beneath the convection zone at which this is half its peak value.  This is shown in Figure~\ref{fig:fwhm}, which shows the radial gradient of angular velocity as a function of radius at both high (left-hand panels) and low (right-hand panels) latitudes at various times.  The dashed lines represent the secondary peak and the solid lines represent half of that peak.  If the tachocline penetration depth is measured as the distance between the base of the convection zone and the solid line, the average tachocline depth is $0.021R_{\odot}$ ($1.46\times 10^{9}$cm) at high latitudes and $0.022R_{\odot}$ ($1.53\times 10^{9}$cm) at low latitudes.\footnote{This provides an upper limit for this way to measure the tachocline depth.  If we had measured from the secondary peak to its half max, the depth would be much smaller.}  Measured this way, the tachocline penetration depths we measure are consistent with helioseismic results. 

\begin{figure}
\centering
%\plotone{f5.pdf}
%\epsscale{0.80}
\includegraphics[width=5in]{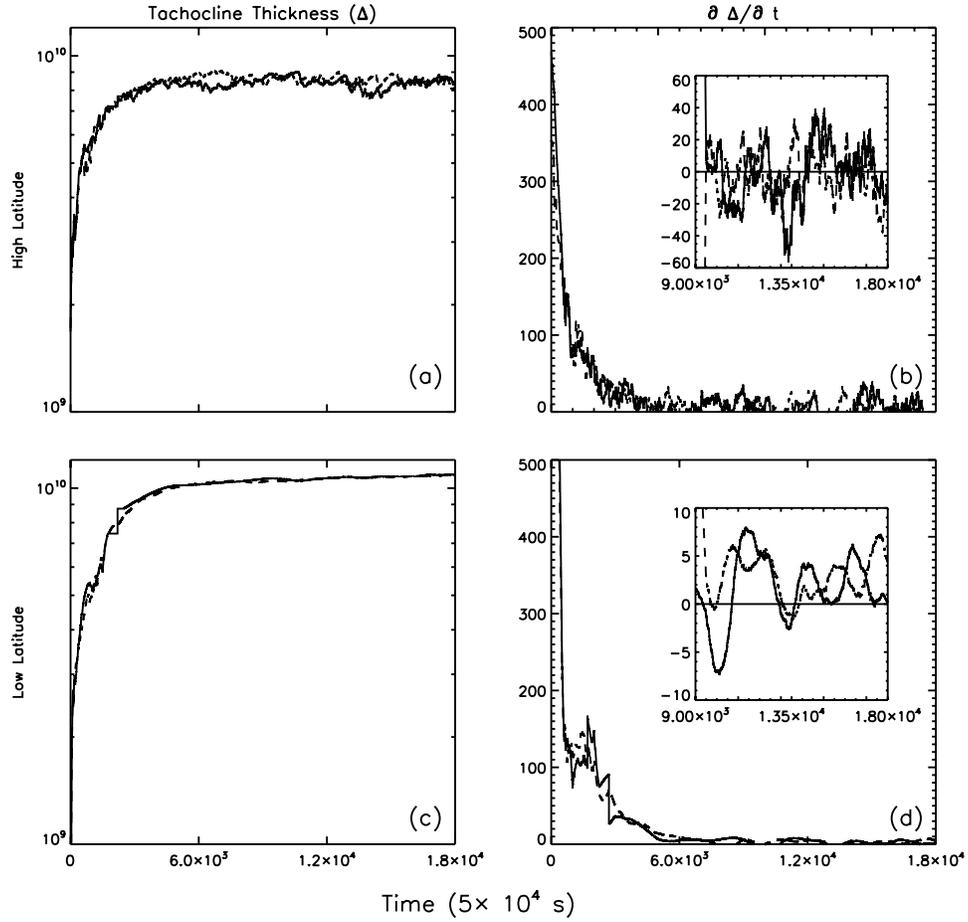}
\caption{Tachocline thickness as a function of time for the non-magnetic model (solid line) and the strong magnetic case (dashed line) in cm.  This thickness is measured as described in the text and is averaged over the polar regions (0$^{\circ}$-30$^{\circ}$, a) and the equatorial regions (60$^{\circ}$-90$^{\circ}$, c).  The time derivative of the tachocline penetration depth is shown in panels (b) and (d) in cm/s.  Polar regions show that the tachocline thickness has stopped increasing.  Lower latitudes still show a slow increase.  The insets in panels (b) and (d) show the time derivative of the tachocline penetration depth over just the last half of the simulation.}
\label{fig:tachheight}
\end{figure}

Another way to define tachocline thickness at a given latitude is the depth beneath the convection zone at which the angular velocity perturbation (initally zero) is X\% of its forced value at that latitude within the convection zone.  We plot this value for X=50\% in Figure~\ref{fig:tachheight}, and note that the general behavior in time at different values of X is the same even if the amplitude is not.  Figure~\ref{fig:tachheight} shows the tachocline thickness, measured this way, as a function of time, averaged over polar regions (0$^{\circ}$-30$^{\circ}$) and equatorial regions (60$^{\circ}$-90$^{\circ}$) for the non-magnetic case (soid line) and the strong field case (dashed line).  Also shown in that figure is the time derivative of the tachocline penetration depth as a function of time (panels (b) and (d), and zoomed in over the last half of the simulation in the insets).  The first obvious feature of these results is that the tachocline thickness increases rapidly, and subsequent evolution is slow.  In fact, in the polar regions the tachocline penetration depth has stopped increasing, as can be seen in Figure~\ref{fig:tachheight}a, as well as ~\ref{fig:tachheight}b, where the time derivative is sometimes negative.  In the equatorial regions the tachocline penetration depth is still increasing, however at a very slow rate, approximately 1-5 cm/s (as seen in the insets of panels b and d of Figure~\ref{fig:tachheight}), which would take longer than the solar age to spread through the entire solar radiative interior.  Measuring the tachocline depth this way we recover tachocline depths of $\approx 8 \times 10^{9} - 10^{10}$cm or $0.12-0.16 R_{\odot}$.

The tachocline widths measured this way are larger and inconsistent with helioseismic results.  However, given the diffusivities used in these simulations a thicker tachocline is not surprising.  Furthermore, the most important feature is that the tachocline thickness has virtually ${\it stopped}$ spreading (particularly at high latitudes).  At high latitudes, the tachocline thickness at the end of the simulation is 97\% that at half the simulation time for the non-magnetic case and 98\% for the magnetic case.  At low latitudes the tachocline thickness is 2\% higher at the end of the simulation compared to half of the simulation time for both magnetic and non-magnetic cases.  

The viscous spread of the tachocline should give a tachocline thickness $\Delta \approx R(t/t_{\nu})^{1/2}$.  After $4\times 10^{7}$s (the time of initial rapid tachocline spread in these simulations) and using the viscous diffusivity at the top of the tachocline, the viscous depth is approximately 0.16 $R_{\odot}$ ($10^{10}$cm), using the viscous diffusivity at the bottom of the tachocline this depth is approximately $0.08 R_{\odot}$ ($5.5\times 10^{9}$cm), in excellent agreement with the tachocline thicknesses seen here\footnote{The initial rapid adiabatic response predicted by \cite{sz92} occurs on a timescale of approximately $2\times 10^{6}$s, shorter than what we see here.  Furthermore, the depth over which it extends is predicted to be approximately $4\times 10^{7}$cm, again shorter than we see here.}.  However, if we were to assume viscous spread over the entire simulation time, we would get tachocline penetration approximately equal to the depth of the radiative interior.  

 If, on the other hand, the subsequent evolution of the tachocline (after the rapid initial spread) obeyed the circulation-enhanced, ``hyper-diffusive'' spreading described in \cite{sz92} in which $\Delta \propto t^{1/4}$, then the tachocline penetration depth would be approximately 1.7 times larger at the end of the simulation than it is at 1/3 of the simulation time.  This is also not seen, clearly indicating that something is preventing both the viscous and the thermal hyper-diffusion spread of the tachocline, {\it even in the non-magnetic case}\footnote{In these simulations the Prandtl number $\nu/\kappa$ in the tachocline is approximately 0.05, but varies with depth.  This is not low enough to ensure that the spread due to thermal hyper-diffusion is larger than that due to viscous diffusion, as in the Sun.}.  

In the following sections, we refer to the ``tachocline'' in these models as defined by the radial gradient of the azimuthal velocity or $0.02R_{\odot}$.  Although we recognize that there are other ways to define the tachocline depth (as studied above) we find that the relevant force balance which stops the tachocline spread occurs in this narrow depth.  We now turn to the question of what is stopping the spread of the tachocline.  

\subsection{The Spiegel \& Zahn Model}

The defining work on the tachocline by \cite{sz92} proposed that the spread of the tachocline could be stopped by an anisotropic turbulence, akin to that seen in the Earth's atmosphere.  This is argued based on the fact that the bulk of the tachocline lies within the radiative interior, where the Brunt-Vaisala frequency is large and likely inhibits substantial radial motion.  Since our non-magnetic model also shows a halt of the tachocline penetration depth this seems a likely candidate.  We can test this proposal by investigating the stresses within the tachocline.  The Reynolds stress can be broken into mean and fluctuating components, such that: 
\begin{equation}
\nabla\cdot({\bf u}u_{\phi})= \nabla\cdot ({\bf u} u_{\phi}')+\nabla\cdot ({\bf u} \bar{u}_{\phi})
\end{equation}
where $\bar{u_{\phi}}$ represents a time average and $u_{\phi}'$ represents the fluctuation, so that $u_{\phi}=\bar{u}_{\phi}+u_{\phi}'$.

\begin{figure}
%\plotone{f6.pdf}
%\epsscale{0.80}
\centering
\includegraphics[width=5in]{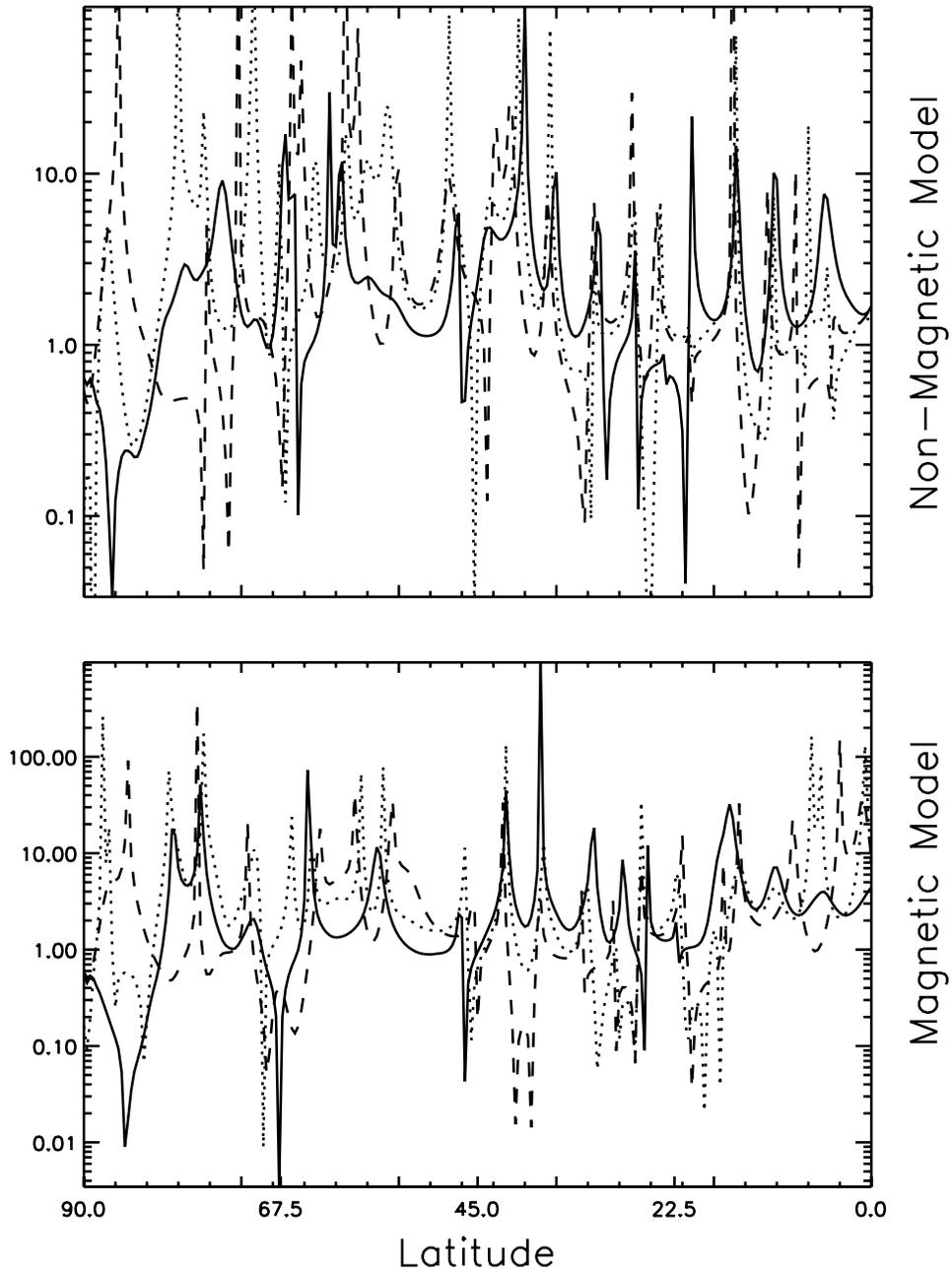}
\caption{Anisotropy of the Reynolds stress, measured as the ratio of the first two terms in Equation 12 to the last two terms in Equation 12. Different line types denote different radii within the tachocline, 0.71$R_{\odot}$ (solid), 0.70$R_{\odot}$ (dashed) and 0.69$R_{\odot}$ (dotted).}
\label{fig:anisotropy}
\end{figure}

The first term on the rhs of Equation 11, the divergence of the Reynolds stress, can be split into horizontal and radial terms and we can ask if there is any anisotropy.  
\begin{eqnarray}
\nabla\cdot({\bf u} u_{\phi}')&=&\frac{1}{r^{2}}\dxdy{(r^{2}u_{\phi}'u_{r})}{r}+\frac{u_{\phi}'u_{r}}{r}\nonumber\\&
&\mbox{}+ \frac{1}{r\sin\theta}\dxdy{(u_{\phi}'u_{\theta}\sin\theta)}{\theta} +\frac{u_{\phi}'u_{\theta}\cot\theta}{r}
\end{eqnarray}
The anisotropy envisioned by \cite{sz92} would then require that the ratio of the radial to latitudinal terms on the rhs of Equation 12 be small.  We show this ratio in Figure~\ref{fig:anisotropy} as a function of latitude at various depths within the tachocline.  In both models and at all radii evaluated in the tachocline there is no detectable anisotropy of the kind envisioned in \cite{sz92}.  This could be due to our lack of resolution, so we can not rule this model out completely, but we can say it is not responsible for the halt of tachocline spread seen here.  In fact, in much of the tachocline these Reynolds stresses are small compared to other terms in Equation 8 (see next section). 

\subsection{Force Balance}

The above discussion then implies that the \cite{sz92} model is not at work in these simulations.  Furthermore, given that the tachocline spread is halted in the non-magnetic cases, we also see that a magnetic field is not necessary in stopping the tachocline spread\footnote{Although see magnetic section for further discussion.}.  This leads to the important question of what is responsible for limiting the tachocline penetration depth.  For an answer we can look directly at the force balance within the tachocline to determine which forces are responsible for stopping this spread.  The relevant equation is the azimuthal component of the momentum equation given by:
\begin{eqnarray}
  \dxdy{}{t}\left(\frac{u_{\phi}}{r \sin\theta}\right)
  &=&-\frac{1}{r\sin\theta}\left(\nabla\cdot\mathbf{u}\bar{u_{\phi}}+\nabla\cdot\mathbf{u}u_{\phi}'\right)
  +\frac{[(\nabla\times{\bf B})\times{\bf B}]_{\phi}}{\bar{\rho}\mu r\sin\theta}\nonumber\\&
  &\mbox{}+\frac{(2{\bf u}\times\Omega)_{\phi}}{r\sin\theta}+\frac{\nu}{r\sin\theta}\nabla^{2}u_{\phi}+\frac{\nu u_{\phi}}{r^{3}\sin^{3}\theta}
\end{eqnarray}
where F has been omitted because there is no forcing in this region and, as in Equation 11, the total azimuthal velocity, $u_{\phi}$ is made up of the mean (time-averaged, $\bar{u_{\phi}}$) plus the fluctuating component, $u_{\phi}'$.

In Figure~\ref{fig:forcebalance} we show the various force terms as a function of latitude, at two different radii within the tachocline, with the top four panels representing the non-magnetic case and the bottom four panels representing the strong magnetic case\footnote{The magnetic stresses are hardly seen because they are substantially lower than the other stresses in the tachocline, see section 3.5.}.  The left-hand panels represent high latitudes, while the right hand panels represent low latitudes.  In the following discussion we will concentrate on the non-magnetic case, but note here that the magnetic case shows the same behavior.  At high latitudes, at the top of the tachocline (top left panel) the total force (denoted by the solid line) is very small.  The force there is dominated by the the Coriolis force (dashed line, defined as the sum of the first and fourth terms on the rhs of Equation 13) and the divergence of the Reynolds stress (dotted line, defined as the second term on the rhs of Equation 13).  The divergence of the Reynolds stress in this region is largely due to convective overshoot and is therefore, unsurprisingly large.  The sign of the Coriolis force at the top of the tachocline is negative indicating its tendency to slow the poles.  However, this Coriolis term is offset entirely by the Reynolds stress and diffusive terms, leading to very little net angular momentum transport there.  

At the bottom of the tachocline, again at high latitudes, the Coriolis term, which is still dominant has changed sign and leads to an ${\it acceleration}$ of the azimuthal flow at high latitudes.  The change of sign of the Coriolis force is because, at the bottom of the tachocline, the Coriolis force is acting on the counter-rotating meridional circulation cell discussed in Section 3.1, which is poleward.  Therefore, this counter rotating cell acts to change the sign of the dominant force in the tachocline, effectively turning off the spread of the tachocline in radius.
  
At lower latitudes, the situation is more complicated.  There is substantially more time and latitudinal variation.  This is because a persistent CRC has not developed at lower latitudes.  We believe that over longer time averages the low-latitude CRC will become more persistent and halt the spread there as well.  This expectation is supported by Figure~\ref{fig:tachheight} which shows the spread of the tachocline nearing zero, even at low latitudes. 

\subsection{Magnetic Models}

As stated above we ran three models: two with a magnetic field and one without.  Although we ran two magnetic cases, as stated previously we only present results from our strong field case, for ease of presentation, but also because the results in the weak field case are virtually identical to the non-magnetic case.  The results for the tachocline penetration depth shown in Figure~\ref{fig:tachheight} and Figure~\ref{fig:fwhm} indicate that the \cite{gm98} model can not be solely responsible for halting the spread of the tachocline.  However, it is also expected that an internal poloidal field once ``connected'' to the differentially rotating convection zone would cause the differential rotation to be spread into the radiative interior on approximately an Alfven crossing time.  In Figure~\ref{fig:fieldevol} we show the evolution of the poloidal (and toroidal) field, where we can clearly see that the poloidal field connects to the convection zone.  Yet, despite this connection we see no additional spread of the tachocline in the magnetic case.  

For our strong field case, with an initial internal field strength of 4kG, assuming no diffusion, nor any turbulent diffusion due to small scale motion, the Alfven velocity varies from a few cm/s at the base of the convection zone to a few hundred cm/s in the deep radiative interior.  After the initial rapid spread of the tachocline as seen in Figure~\ref{fig:tachheight}, that is at $\approx 0.6R_{\odot}$, the Alfven velocity ($v_{a}$) is approximately 100 cm/s.  Therefore, in the subsequent $10^{8}$s after the initial spread of differential rotation, one would (very simply) expect an additional spread of $v_{a}t\approx 10^{10}$cm if that spread were being mediated by a magnetic field.  However, we see no spread of this magnitude.  Indeed, we see very little effect of the magnetic field at all.

\begin{figure}
%\plotone{fig9.pdf}
%\epsscale{0.80}
\centering
\includegraphics[width=5in]{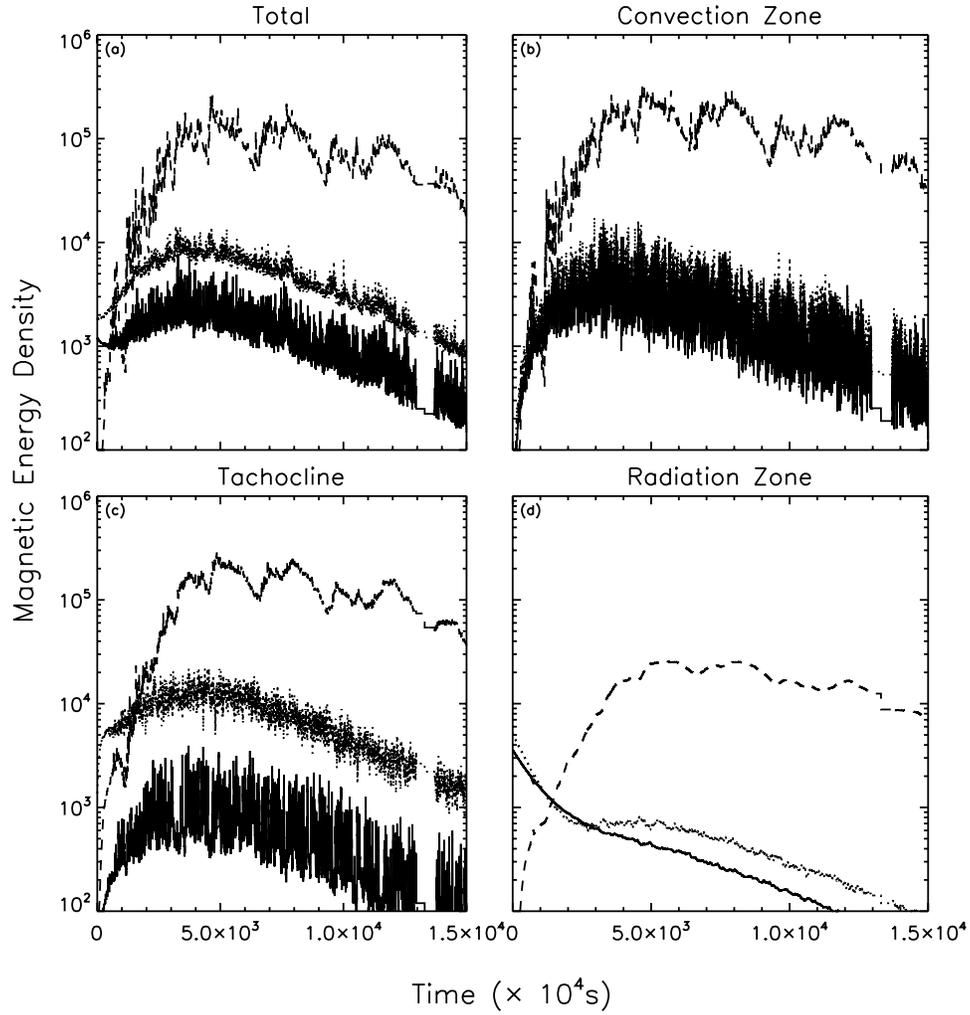}
\caption{Magnetic Energy as a function of time for the three different field components; radial field (solid line), latitudinal field (dotted line) and toroidal field (dashed line).  The total integrated energy for the entire domain are shown in (a), the convection zone (b), the tachocline (c) and the radiative interior (d).  One can clearly see an initial increase in all components of the magnetic field energy in the convection zone and tachocline followed by exponential decay.  The toroidal field is amplified substantially and shows oscillatory behavior where the toroidal energy grows, decays and repeats.  In the raditiative interior, the field predominantly decays, except the toroidal field.}
\label{fig:magen}
\end{figure}

There are (at least) three reasons for this (that this author can think of).  The first, and most obvious, is that these simulations have not run long enough to see the additional effects of the magnetic field.  The estimate above, is just that, a very crude estimate.  Because we wanted our magnetic Prandtl number to be less than one, our magnetic diffusivity is rather large.  Compounding this problem we have no dynamo and therefore, the field strength and Alfven velocity are decreasing (after an initial rise, see Figure~\ref{fig:magen}).  Estimating the decay of the magnetic field strength using the imposed diffusivity, the amplitude of the magnetic field (and hence, Alfven velocity) would have dropped by approximately 30\% and hence, the spread might have only amounted to $7\times 10^{9}cm$, but this would still be measurable.  The picture is more complicated though, because the evolution of the magnetic field is not governed solely by diffusion.  As can be seen in Figure~\ref{fig:magen}, initially the magnetic energy increases due to differential rotation and convection and only later does exponential decay set in.  Depending on the field component, the field decay requires nearly half of the simulation time to decay back to its original amplitude.  Therefore, estimating how far the differential rotation should have spread if mediated by the magnetic field is not straightforward, given the complications of a time and space dependent field.  In light of the uncertainty in estimating the timescale of tachcocline spread mediated by magnetic field, it is wise to at least investigate the ${\it trend}$ of the magnetic stresses.    

\subsubsection{The Lorentz Force in the Tachocline}

The scenario in which a magnetic field aids the spread of differential rotation into the radiative interior depends on a few factors.  First, the field must be sufficiently ``connected'' to the convection zone.  Second, the Lorentz force must be larger than other forces in the region (at least on long time averages).  And finally, the configuration of $B_{\phi}, B_{\theta}$ and $B_{r}$ must be specific.  

\begin{figure}
%\plotone{fig7.pdf}
%\epsscale{0.80}
\centering
\includegraphics[width=5in]{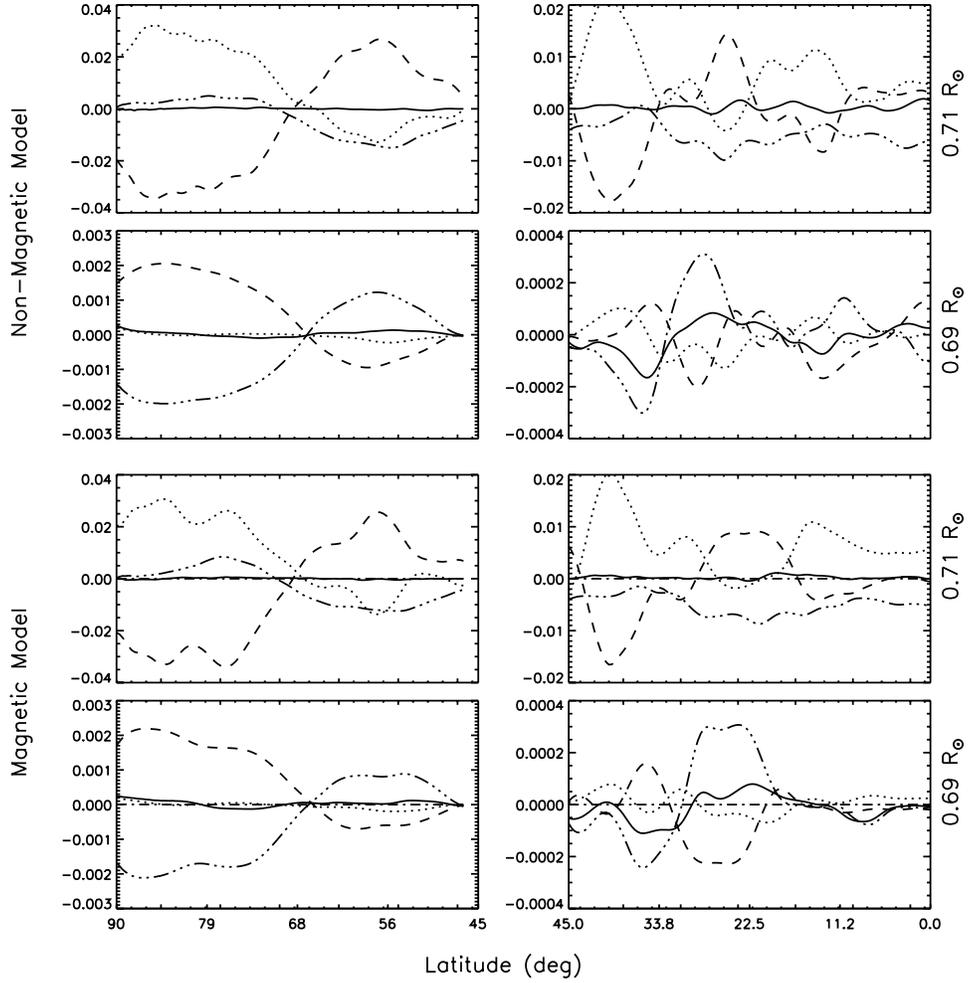}
\caption{Forces within the tachocline in the non-magnetic model (left-hand panels) and in the magnetic model (right hand panels).  Three different radii within the tachocline are shown.  Solid line represents the total force, i.e. the lhs of Equation 13 (all forces are shown in units of $gm /(cm s)^{2}$).  Dashed line represents the Coriolis force (the sum of the first and fourth terms on the rhs of Equation 13).  Dotted line represents the Reynolds stress (second term on rhs of Equation 13) and triple-dot dashed line represents the diffusive terms (the sum of the fifth and sixth terms on rhs of Equation 13).  In the magnetic model the Maxwell stresses (third term on rhs of Equation 13) are also shown (although the appear as zero).}
\label{fig:forcebalance}
\end{figure}

As we have seen in Figure~\ref{fig:fieldevol}, the initial poloidal field indeed becomes ``connected'' to the convection zone and is not confined by the meridional circulation.  Given this clear connection one would expect the differential rotation of the convection zone to be rapidly imprinted into the radiative interior.  However, a magnetically-aided tachocline spread depends on the Lorentz force being dominant or comparable to the other stresses in the azimuthal momentum equation (at least on long time averages).  This depends not only on the strength of the magnetic field in this region but also on the correlation of these fluctuating fields.  In Figure~\ref{fig:forcebalance} we showed the various forces contributing to the maintenance of the differential rotation.  In that figure the Lorentz forces are seen as zero because they are substantially smaller than the other forces shown.  In Figure ~\ref{fig:magen} we show the volume integrated magnetic energy in the various regions of the simulation domain (total: (a), convection zone: (b), tachocline: (c) and radiation zone: (d)).  We see there that the magnetic energy is initially amplified in the convection zone and tachocline by the action of differential rotation and convection.  In particular, the energy in the toroidal field (initially zero) is amplified to the extent that it becomes the dominant field component (dashed line).  Simply estimating the Lorentz force given these amplitudes one would expect it to be comparable to the divergence of the Reynolds stress in the tachocline (where the velocities are approximately $10-10^{3}$ cm/s), but as shown in Figure~\ref{fig:forcebalance}, the Lorentz forces are quite small, indicating that the field fluctuations are correlated in such a way that they do not lead to a substantial Lorentz force.  This can possibly be explained by considering that wherever $B_{\phi}$ is large it is so because differential rotation and/or convection have acted on $B_{r}$ or $B_{\theta}$ to generate it, so that wherever $B_{\phi}$ is large $B_{r}$ and $B_{\theta}$ are not.  Therefore, while the field clearly ``connects'' to the tachocline and convection zone (see Figure~\ref{fig:fieldevol}), the Lorentz force in the tachocline is small and therefore, little angular momentum is transported by magnetic terms.  This is the second possible reason the magnetic field has no effect on the propagation of differential rotation into the radiative interior.  

The idea that a magnetic field could be connected to the convection zone without transferring substantial angular momentum was suggested by \cite{gr07}.  Field lines connecting to the convection zone are made small scale by convection and turbulence, thus imposing an enhanced diffusivity on the magnetic field.  This enhanced diffusivity could act to disconnect the field in the convection zone (in the sense of angular momentum transfer) from the radiative interior.  

\subsubsection{Sense of the Lorentz Force}

\begin{figure}
%\plotone{fig8.pdf}
%\epsscale{0.80}
\centering
\includegraphics[width=5in]{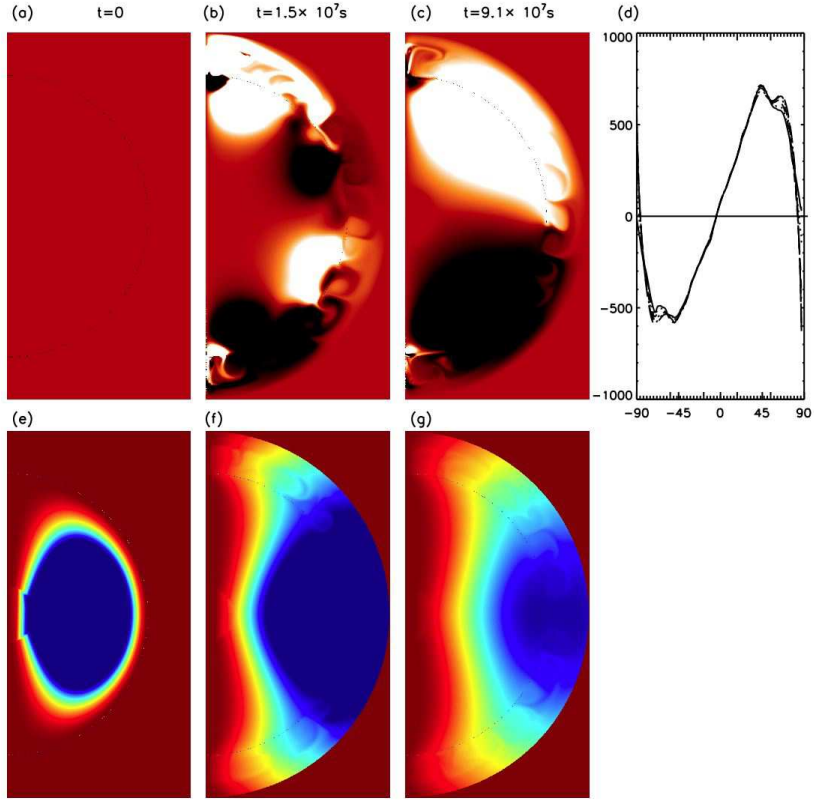}
\caption{Evolution of toroidal (top panels) and poloidal (bottom panels) field.  The dark line represents the base of the convection zone.  {\it Toroidal field evolution}:  Initially, there is no toroidal field (a).  Differential rotation acts on the poloidal field to produce a toroidal field which initially changes sign at mid-latitudes (b).  This configuration then becomes unstable and leaves behind a single dominant sign in each hemisphere (c).  The time averaged toroidal field is shown in (d).  {\it Poloidal field evolution}.  After a short time the poloidal field diffuses into the convection zone (f), aided by convective overshoot and dredge-up, thus connecting the radiative interior to the convection zone. }
\label{fig:fieldevol}
\end{figure}

Finally, the spread of differential rotation by an internal magnetic field depends on the configuration of the magnetic field generated at the convective-radiative interface.  Simply looking at the azimuthal component of the induction equation (neglecting diffusion): 

\begin{eqnarray}
\dxdy{B_{\phi}}{t}&=&rB_{r}\dxdy{}{r}(\frac{u_{\phi}}{r})+\frac{sin\theta B_{\theta}}{r}\dxdy{}{\theta}(\frac{u_{\phi}}{sin\theta})\nonumber\\&
&\mbox{}-ru_{r}\dxdy{}{r}(\frac{B_{\phi}}{r})-\frac{sin\theta u_{\theta}}{r}\dxdy{}{\theta}(\frac{B_{\phi}}{sin\theta}).
\end{eqnarray}
one can see that the in the tachocline, where the radial gradient of differential rotation is initially dominant in inducing toroidal field (first term in Equation 14), a toroidal field which switches sign at mid-latitudes is induced (Figure~\ref{fig:fieldevol}b).  Considering the azimuthal component of the Lorentz force:
\begin{equation}
[(\nabla\times\mathbf{B})\times{\mathbf{B}}]_{\phi}=\frac{B_{\theta}}{r\sin{\theta}}\dxdy{(B_{\phi}\sin\theta)}{\theta}-\frac{B_{r}}{r}\dxdy{(rB_{\phi})}{r}
\end{equation}
a toroidal field which switches sign at mid-latitudes, coupled with a latitudinal field which is positive generates a Lorentz force which promotes decelerating angular velocity at high latitudes and accelerating angular velocity at low latitudes, thereby promoting the spread of the differential rotation of the convection zone into the radiative interior.\footnote{The same sense of Lorentz force is found if the induction of toroidal field is dominated by the latitudinal gradient of angular velocity, because the gradient changes sign with latitude.}

\begin{figure}
%\plotone{f11.pdf}
%\epsscale{0.80}
\centering
\includegraphics[width=5in]{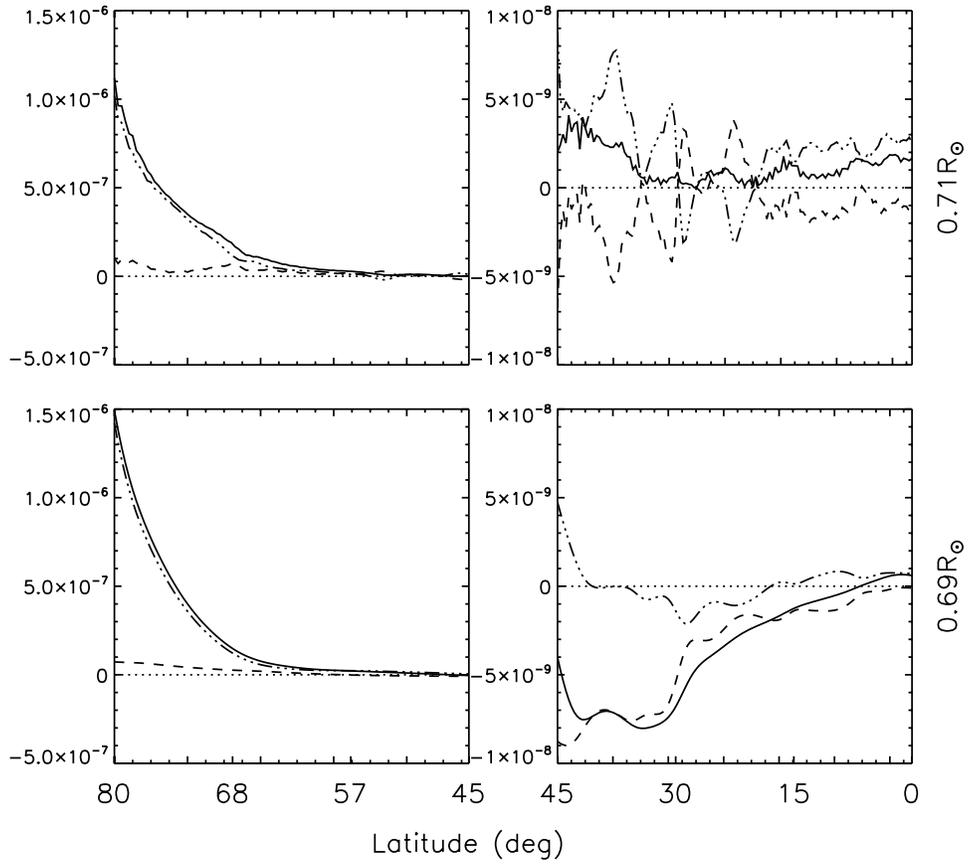}
\caption{Lorentz forces within the tachocline.  Solid line represents the total Lorentz force (rhs of Equation 15), dashed line represents the second term on the rhs of Equation 15 and the triple-dot dashed line represents the first term on the rhs of Equation 15.  The dotted lines denote zero.  The left two panels show high latitudes, while the right two panels show only the lower latitudes.  The top two panels are at the top of the tachocline, $0.71R_{\odot}$, while the bottom two panels show the components of the Lorentz force at the bottom of the tachocline at $0.69R_{\odot}$.}
\label{fig:lorfor}
\end{figure}

In these simulations we indeed find that the radial gradient of differential rotation dominates the initial induction of the toroidal field, leading to a toroidal field which switches sign at mid-latitude (Figure ~\ref{fig:fieldevol}b).  Later in the simulation, however, the latitudinal component of differential rotation becomes dominant.  This leads to a toroidal field configuration which has one sign toroidal field in the northern hemisphere (positive) and one sign in the southern hemisphere (negative), Figure ~\ref{fig:fieldevol}c.  The amplitude of the toroidal field peaks at around mid-latitudes (see Figure~\ref{fig:fieldevol}d), so that the latitudinal gradient of toroidal field changes sign at mid-latitudes.  Furthermore, as the field diffuses out of the radiative interior and connects with the convection zone, the sign of the latitudinal magnetic field changes at high latitudes.  Whereas initially the latitudinal field was positive everywhere in the tachocline, later in the simulation the latitudinal field is negative at high latitudes and positive at low latitudes.  Combining the toroidal field configuration seen in Figure ~\ref{fig:fieldevol}d, with positive latitudinal field at low latitudes and negative latitudinal field at high latitudes, one can then simply see that this field configuration gives ${\it acceleration}$ at high latitudes and ${\it deceleration}$ at low latitudes, as seen in Figure ~\ref{fig:lorfor}.  This field configuration would then act to stop tachocline spread (if Lorentz forces were dominant over long timescales).  

Of course, it is not clear that such a field configuration would develop in the Sun.  But this highly simplified simulation indicates that the magnetic field configuration and associated Lorentz force in a tachocline that is bounded on top by convection and fed from below by a space and time-dependent magnetic field is quite complicated and possibly not well described without considering instabilities and poloidal field evolution.  

\subsection{Weaknesses}

Clearly, this model has several weaknesses.  The first is that it is essentially two-dimensional and there is some possibility that two-dimensional convection conspires to limit tachocline spread.  It is impossible to test this until three-dimensional simulations, using similar parameters as those described here are conducted.  Such simulations are likely not too far off.  

The second major limitation of these simulations is that the diffusivities are too large.  This prevents us from directly testing the case in which the thermal hyper-diffusion described by \cite{sz92} dominates the viscous diffusion.  It also prevents us from getting the force balance in the tachocline exactly right.  For instance,  our meridional circulation appears to be only about 45 times to large, leading to Reynolds stresses and Coriolis forces in the tachocline which are larger than expected.  However, this should be more offset again by the larger than solar viscous diffusivity.  In the end, the viscous forces are not dominant in the tachocline in these simulations and that is the first order goal.

Finally, because our timestep is severely limited by gravity waves and convection we are not able to run this simulation very long.  We have only run these simultations approximately 20\% of a viscous diffusion time or 40\% a magnetic diffusion time.  Over this time we would have expected more tachocline spread and we are at least able to at least the trend of the stresses.  The following discussion then proceeds with these weaknesses in mind. 
   
\section{Discussion}

These results represent the first attempt at coupling the convective and radiative interior of the Sun in order to understand the angular momentum evolution of the radiative interior.  Although troubled by computational limitations (as all simulations are), there are still some very interesting and important results to be gleaned from this model.  First, we find that a purely hydrodynamical model of the solar interior could prevent the spread of the tachocline by setting up a counter-rotating cell that straddles the base of the convection zone/top of the tachocline.  The Coriolis force acting on this CRC acts to accelerate the flow at high latitude, thus shutting off the spread of the tachocline.

In the magnetic models the meridional circulation appears unable to prevent the field from connecting to the convection zone; both by diffusion and dredge-up by convective plumes.  However, despite being connected to the convection zone very little angular momentum is transported between the two regions by magnetic stresses.  This is because the magnetic stresses in the tachocline are small.  Despite a rather large toroidal field growing in the tachocline due to differential rotation, the radial and latitudinal field there are quite small and the Maxwell stress resulting from the correlations of these field components are similarly small.  Therefore, while it does not appear necessary for a magnetic field to prevent the spread of the tachocline, we also find that such a internal poloidal field does not promote differential rotation if it connects to the convection zone.  

\bibliographystyle{apj}
\bibliography{solarrot}

\acknowledgments
We are grateful to G. Glatzmaier, K.B. MacGregor, M. Rempel and P. Garaud for helpful discussions.  The manuscript was substantially improved due to very helpful and insightful suggestions by an anonymous referee.  Support for this research was provided by a NASA grant NNG06GD44G.  T. Rogers is supported by an NSF ATM Faculty Position in Solar physics under award number 0457631.  Computing resources were provided by NAS at NASA Ames.  
\email{aastex-help@aas.org}.

\clearpage

\clearpage

\clearpage

\clearpage

\clearpage

\clearpage
\clearpage
\clearpage

\clearpage

\clearpage

\clearpage

\clearpage

\end{document}